\documentclass[12pt]{article}
\usepackage{epsfig}
\usepackage{graphicx}
\usepackage{a4}
\usepackage{latexsym}
\usepackage{cite}

\usepackage{color}
\usepackage{colordvi}

\usepackage{pslatex}
\usepackage[latin1]{inputenc}
\usepackage[T1]{fontenc}

\newcommand{\as}{\alpha_{\rm s}}

\def\MSbar{\overline{\mathrm{MS}}}

\def\nf{{n^{}_{\! f}}}
\def\nl{{n^{}_{\! l}}}

\def\LOGmu2Om2{\log\left({\mu^2\over m^2}\right)}

\begin{document}
\setlength{\parskip}{0.2cm} \setlength{\baselineskip}{0.55cm}

\begin{titlepage}
\noindent \vspace{2.0cm}
\begin{center}

\LARGE {\bf On the Soft-Gluon Resummation in Top Quark Pair Production at
  Hadron Colliders} \\
\vspace{2.6cm}
\large
M. Czakon$^{a,b}$ and A. Mitov$^{c}$ \\
\vspace{1.4cm}
\normalsize {\it
$^{a}$Institute of Nuclear Physics, NCSR ``DEMOKRITOS'',
15310 Athens, Greece \\[.5cm]
$^{b}$Fachbereich C, Bergische Universit\"at Wuppertal, D-42097, Wuppertal,
Germany \\[.5cm]
$^{c}$C. N. Yang Institute for Theoretical Physics, Stony Brook
University, Stony Brook, NY 11794, USA} \vfill
\large {\bf Abstract}
\vspace{-0.2cm}
\end{center}

We uncover a contribution to the NLO/NLL threshold resummed total cross
section for top quark pair production at hadron colliders, which has not been
taken into account in earlier literature. We derive this contribution -- the
difference between the singlet and octet hard (matching) coefficients -- in
exact analytic form. The numerical impact of our findings on the Sudakov
resummed cross section turns out to be large, and comparable in size to the
current estimates for the theoretical uncertainty of the total cross
section. A rough estimate points toward a few percent decrease of the
latter at the LHC.

\vspace{3.0cm}
\end{titlepage}

%
%
\section{Introduction}
\label{sec:intro}

Improving the theoretical accuracy of the total cross section for
top quark pair production is a major goal, given the importance of
top physics, the excellent data-taking ability of the Tevatron and
the imminent start of the LHC. Yet, relatively few theoretical
calculations for this observable have been done so far. The most
important input are the next to leading (NLO) numerical calculations
of \cite{Nason:1987xz,Beenakker:1988bq,Bernreuther:2004jv}, where
the NLO correction with accuracy better than 1\% was derived. The
first exact calculation of this observable \cite{Czakon:2008ii}
confirmed these results and their estimates of the numerical
uncertainties. By demonstrating the appearance of a priori
unexpected analytic structures, that work also clarified why
theoretical progress in top production was hampered for so long.
While work towards the derivation of the NNLO corrections to the
total top quark pair production cross section is underway
\cite{Czakon:2007wk,Czakon:2007ej,Czakon:2008zk,Korner:2008bn,Kniehl:2008fd,Bonciani:2008az,Anastasiou:2008vd}
more theoretical effort will be needed before the NNLO result
becomes available for phenomenological analysis.

In view of the lack of improved fixed order calculations in the last ten years
or so, the only other source of refinement in the theoretical predictions for
the total top quark pair production cross section was based on the so-called
soft-gluon resummation
\cite{Sterman:1986aj,Catani:1989ne,Catani:1990rp,Laenen:1993xr,Catani:1996yz,
  Bonciani:1998vc} (related phenomenological analyses can be found in
\cite{Moch:2008qy,Cacciari:2008zb,Kidonakis:2008mu}).  The basic idea is that
our ability to predict certain logarithmic terms to all orders in the strong
coupling expansion provides insight into the behavior of the cross section at
higher orders.  That, in turn, implies better control over the associated
theoretical uncertainties.

Heavy flavor pair production at hadron colliders plays a prominent
role in the global soft-gluon resummation program. The reason is
that at least four hard partons are involved in the underlying
scattering process. As it is well known (see, for example,
Ref.~\cite{Bonciani:2003nt}) in the case of scattering of four or
more colored partons the color algebra cannot be made trivial. This,
in turn, spoils the simple exponentiation picture familiar from
processes like Drell-Yan and Deep Inelastic Scattering.

As was first established in Ref.~\cite{Kidonakis:1997gm}, the one-loop soft
anomalous dimension matrix that controls the non-trivial 
soft-gluon correlations in this process diagonalizes in the
singlet/octet basis in the kinematical configuration in question.
This is a very important result that is the basis for the simplified
exponentiation of the next-to-leading soft logarithms (NLL) in this
process.

Utilizing the above results, the exponentiation formula for the
total inclusive top quark pair production at hadron colliders beyond
the leading logs was given in Ref.~\cite{Bonciani:1998vc}. The
singlet/octet diagonalization mentioned above implies that the soft
function $\Delta$ is just a sum $\Delta = \Delta_{\bf 1} +
\Delta_{\bf 8}$ of two standard Sudakov-type exponents, which
separately describe the exponentiation of the singlet (resp. octet)
color channels. Each one of the two exponents is controlled by its
own set of anomalous dimensions
\begin{equation}
\label{eq:Sudakov-exp} \ln \Delta_{ij,{\bf I}}(N) = \int_0^1 dz
{z^{N-1}-1 \over 1-z} \left( \int_{\mu^2}^{4m^2(1-z)^2} {dq^2\over
q^2} A_{ij}(\as(q^2)) + D_{ij,\bf I}( \as( 4m^2(1-z)^2) \right) \, .
\end{equation}
The indices $ij$ refer to the partons in the initial state: $ij =
(q{\bar q}, gg)$, ${\bf I}={\bf 1,8}$ and $N$ is the Mellin moment
dual to the kinematical variable $\rho = 4m^2/s$ (with $s$ being the
partonic invariant mass). The Mellin transform is defined as $f(N) =
\int_0^1 \rho^{N-1} f(\rho) d\rho$. The anomalous dimensions $A_{ij}
= A_i+A_j$ describe soft-collinear initial state radiation. They
have a standard expansion in powers of the running strong coupling,
see e.g. Ref.~\cite{Bonciani:1998vc}, and are known in QCD through
three loops \cite{Moch:2004pa,Vogt:2004mw}. Contrary to $A_{ij}$,
the anomalous dimensions $D_{ij, \bf I}$ control wide angle soft
radiation and depend both on the initial and final states. They are
a priori unknown, but one linear combination of $D_{ij,\bf 1}$ and
$D_{ij,\bf 8}$ can be fixed from existing results for the threshold
expansion of the total cross-section
\cite{Nason:1987xz,Beenakker:1988bq,Czakon:2008ii} at NLO. To fix
uniquely both anomalous dimensions at the same order of perturbative
expansion, the authors of  Ref.~\cite{Bonciani:1998vc} used
heuristic arguments, namely, that the Sudakov exponent for the color
singlet channel is identical to the one known from processes like
Drell-Yan and Higgs production.  With the help of an explicit
calculation, in this paper we are able to directly confirm that
assumption through NLO/NLL.

Besides the soft Sudakov exponents $\Delta_{ij,{\bf I}}$, color
dependence is also present in the so-called hard coefficients that
we discuss next. In the singlet/octet basis one expresses the
Sudakov total cross-section for top quark pair production at hadron
colliders within NLL accuracy as
\begin{equation}
\label{eq:sigma-tot} \sigma^{\rm TOT}_{ij}(N) = \sigma_{ij,{\bf
1}}(N) + \sigma_{ij,{\bf 8}}(N) \, ,
\end{equation}
where the two terms are given by
\begin{equation}
\label{eq:sigma-one-or-eight} \sigma_{ij,{\bf I}}(N) = \sigma^{\rm
Born}_{ij,{\bf I}}(N)~\sigma^{\rm H}_{ij,{\bf I}}~\Delta_{ij,{\bf
I}}(N) \, .
\end{equation}

The hard coefficients $\sigma^{\rm H}_{ij,{\bf I}}$ are process
dependent and once the leading order correction $\sigma^{\rm
Born}_{ij,{\bf I}}(N)$ has been factored out, they contain only
$N$-independent constant terms. As usual, these coefficients are
extracted from a fixed order calculation. Their derivation through
NLO is the main goal of this article.

In Eq.~(\ref{eq:sigma-one-or-eight}) we omitted contributions from
Coulomb terms, i.e. terms that in $\rho$-space behave as $\sim
\as^n/\beta^k$, where $\beta = \sqrt{1-\rho}$ is the small velocity
of the quark pair. The Coulomb terms represent an effect distinct
from the soft gluon logs considered in this article. These
corrections have been analyzed in
Ref.~\cite{Catani:1996dj,Bonciani:1998vc,Hagiwara:2008df} with the
conclusion that they have an impact only in the immediate vicinity
of the threshold. We refer to these references for further details.

Next we explain the origin of the color index $I$ in the hard
functions appearing in Eq.~(\ref{eq:sigma-one-or-eight}). The
easiest way to see why it should be present is to recall the basic
factorization property of gauge amplitudes
\cite{Akhoury:1978vq,Sen:1982bt}. Keeping explicit only information
about the color, the factorization relation for any $n$-particle
amplitude $M$ reads
\begin{equation}
\label{eq:factorization-amplitude} M_I = J\cdot S_{IJ} \cdot H_J \,
.
\end{equation}

In the equation above $S_{IJ}$ is the soft function mentioned above,
$I,J$ are color indices and $H$ is the so-called hard function,
which is finite. While the structure of the color diagonal jet
function $J$ and the soft function $S$ can be made quite transparent
based on general process-independent arguments
\cite{Kidonakis:1998nf,Sterman:2002qn,hep-ph/0607309,MS1}, the form
of the process dependent hard function $H_J$ can only be obtained
from a direct, process specific calculation. The matrix structure in
Eq.~(\ref{eq:factorization-amplitude}) naturally translates into
differential or fully integrated over the phase space
cross sections. An explicit example for that procedure can be found
in Ref.~\cite{Kidonakis:2001nj}.

The color dependence of the hard coefficients $\sigma^{\rm
H}_{ij,{\bf I}}$ was not available to any of the previous studies of
soft-gluon resummation for the total inclusive cross section in
hadronic collisions. The main goal of this article is to complete
this gap in the literature by deriving the {\it exact} coefficients
from a dedicated fixed order calculation, thus allowing a consistent
NLO/NLL calculation and soft gluon resummation for this observable.
It is also a prerequisite for any attempt for going beyond the
current NLL accuracy level.

In the original Ref.~\cite{Bonciani:1998vc} these coefficients were
approximated with the numerically known, color averaged coefficient
taken from the calculations of
Refs.~\cite{Nason:1987xz,Beenakker:1988bq}. Such an approximation is
formally correct if one restricts oneself only to the resummation of
the NLL soft logs since, as far as the towers of logs are concerned,
these matching coefficients contribute starting from NNLL. On the
other side, such an approximate choice is also correct to NLO, since
by construction it reproduces the fixed order NLO results for the
color summed cross-section in Eq.(\ref{eq:sigma-tot}). Nevertheless,
one expects that the specific choice does have a numerical impact on
the resummed cross-section. This is easy to see with the help of the
following argument: the total Sudakov cross section is a linear
combination of two all-order exponents (see Eq.(\ref{eq:sigma-tot}))
with coefficients proportional to the hard coefficients $\sigma^{\rm
H}_{ij,{\bf I}}$. Therefore, a modification of the two coefficients
in such a way that their color-averaged contribution is kept fixed,
results in a change of the weight these two exponents
(singlet/octet) carry.

The effect of this modification should be much less pronounced at
the Tevatron compared to the LHC since there the main production
mechanism for top-pair production is through light quark pair
annihilation. It is known that for the $q{\bar q}$ production
through NLO/NLL only color octet contributes, and in this case the
hard coefficient has always been known with high accuracy.

One final comment regarding the hard coefficients $\sigma^{\rm
H}_{gg,{\bf I}}$. As was established in Ref.~\cite{Czakon:2008ii}
and also investigated in Ref.~\cite{Hagiwara:2008df}, the
``constant'' term in the color averaged $gg$ cross-section extracted
from \cite{Nason:1987xz} differs by around 7\% from the exact value.
It is reasonable to suspect that this correction might be quite
sizable. We detail our findings in the next section.

%
%
\section{Results}
\label{sec:results}

\begin{figure}[t]
  \begin{center}
    \epsfig{file=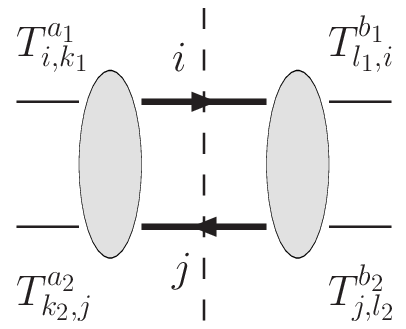, width=5cm}
  \end{center}
  \caption{\it \label{fig:proj} Color indices in a cut graph for top quark
    pair production, necessary to define the color projection onto
    singlet/octet states Eq.~(\ref{eq:proj}). The dashed line represents the
    cut through the top line with color index $i$, and the anti-top line with
    color index $j$.}
\end{figure}

To derive the hard coefficients $\sigma^{\rm H}_{ij,{\bf I}}$ we
follow the technique described in Ref.~\cite{Czakon:2008ii}. The
only required modification consists in the need to insert suitable
color projection operators in order to separate the contributions
where the heavy pair is in singlet/octet state. To this end we define the
singlet state as
\begin{equation}
| {\mathbf 1}, \; i, j \rangle = \frac{1}{\sqrt{N}} \delta_{ij} | {\mathbf 1}
\rangle \; ,
\end{equation}
where $i$ and $j$ are the color indices of the quark and anti-quark
respectively. The projection onto the singlet state $|{\mathbf 1}
\rangle$, can now be performed with the simple replacement
\begin{equation}
\label{eq:proj} T^{a_1}_{i,k_1} T^{a_2}_{k_2,j} T^{b_1}_{l_1,i}
T^{b_2}_{j,l_2} = \frac{1}{N_c} T^{a_1}_{i,k_1} T^{a_2}_{k_2,i}
T^{b_1}_{l_1,j} T^{b_2}_{j,l_2} \; ,
\end{equation}
where the color indices can be understood with the help of
Fig.~\ref{fig:proj}. The remaining contribution is simply attributed to the
color octet state.

It is interesting to note that the color separation is well defined
only in the vicinity of the threshold. Further away, both
contributions become separately divergent. This is not really
surprising, since the octet state will have a tendency to attract
radiated gluons and hadronize into a singlet. The necessity to
combine both singlet and octet contributions in order to obtain
finite cross sections is only visible starting from ${\cal
O}\left(\beta^3\right)$, and does not affect our discussion of
soft-gluon effects.

Once the color separation has been accomplished, the result needs to
be expanded around threshold. The resulting expressions (keeping
also the Coulomb terms) read
\begin{eqnarray}
\sigma_{q{\bar q},{\bf 1}}(\beta) &=& \as^3 \times {\cal O}(\beta^3)
\, , \label{eq:qqbar-1-beta}\\
\nonumber\\
\sigma_{q{\bar q},{\bf 8}}(\beta) &=& \sigma^{\rm Born}_{q{\bar
q},{\bf 8}}(\beta)~\left\{ 1+ {\as\over \pi}\left[
\left(C_F-{C_A\over 2}\right){\pi^2\over 2\beta} + 8C_F\log^2 \beta
\right.\right.
\nonumber\\
&+& \Bigg(  C_F (-16 + 24 \log 2)  -2 C_A \Bigg) \log\beta + C_F
\left(8 - {\pi^2\over 3} - 21 \log 2 + 16 \log^2 2\right)
\nonumber\\
&+& C_A \left( {77\over 9} - {\pi^2\over 4} - 5 \log 2 \right) + \nl
\left( -{5\over 9} + {2 \log 2\over 3} \right) - {8\over 9} +
\LOGmu2Om2 \Bigg( - 4 C_F \log\beta
\nonumber\\
&+&\left.\left.  C_F \left({5\over 2} - 4 \log 2\Bigg)+ {11\over 6}
C_A - {\nl + 1\over
3} \right) + {\cal O}(\beta) \right] + {\cal O}(\as^2)\right\} \, , \label{eq:qqbar-8-beta}\\
\nonumber\\
\sigma_{gg,{\bf 1}}(\beta) &=& \sigma^{\rm Born}_{gg,{\bf
1}}(\beta)~\left\{ 1 + {\as\over \pi}\left[ C_F{\pi^2\over 2\beta} +
8C_A\log^2 \beta + C_A \left(  -16 + 24 \log 2 \right)
\log\beta\right.\right.
\nonumber\\
&+& C_F \left(-5 + {\pi^2\over 4}\right) + C_A \left( 17 -
{7\pi^2\over 12} - 24 \log 2 +16\log^2 2\right)
\nonumber\\
&+&\left.\left. \LOGmu2Om2 \left( - 4 C_A \log\beta + C_A \left(4 -
4 \log 2\right) - {1\over 3} \right) + {\cal O}(\beta) \right] +
{\cal O}(\as^2) \right\} \, ,
\label{eq:gg-1-beta}\\
\nonumber\\
\sigma_{gg,{\bf 8}}(\beta) &=& \sigma^{\rm Born}_{gg,{\bf
8}}(\beta)~\left\{ 1 + {\as\over \pi}\left[ \left(C_F-{C_A\over
2}\right){\pi^2\over 2\beta} + 8C_A\log^2 \beta + C_A \left(  -18 +
24 \log 2 \right) \log\beta\right.\right.
\nonumber\\
&+& C_F \left(-5 + {\pi^2\over 4}\right) + C_A \left( 21 -
{17\pi^2\over 24} - 26 \log 2 +16\log^2 2\right)
\nonumber\\
&+&\left.\left. \LOGmu2Om2 \left( - 4 C_A \log\beta + C_A \left(4 -
4 \log 2\right) - {1\over 3} \right) + {\cal O}(\beta) \right] +
{\cal O}(\as^2) \right\} \, , \label{eq:gg-8-beta}
\end{eqnarray}
where
\begin{eqnarray}
\sigma^{\rm Born}_{q{\bar q},{\bf 1}}(\beta) &=& 0 \, , \label{eq:born-qq-1-beta}\\
\sigma^{\rm Born}_{q{\bar q},{\bf 8}}(\beta) &=& {\pi\as^2 \over 8
m^2} ~ {\left( N_c^2-1\right)\over N_c^2}~\beta + {\cal O}(\beta^3)\, ,\label{eq:born-qq-8-beta}\\
\sigma^{\rm Born}_{gg,{\bf 1}}(\beta) &=& {\pi\as^2 \over 4
m^2} ~ {1\over N_c\left(N_c^2-1\right)}~\beta + {\cal O}(\beta^3) \, , \label{eq:born-gg-1-beta}\\
\sigma^{\rm Born}_{gg,{\bf 8}}(\beta) &=& {\pi\as^2 \over 8 m^2} ~
{N_c^2-4\over N_c \left(N_c^2-1\right)}~\beta + {\cal O}(\beta^3) \,
. \label{eq:born-gg-8-beta}
\end{eqnarray}
The coupling $\as$ is the renormalized $\MSbar$ coupling evaluated
at scale $\mu^2$ and running with $\nf = \nl+1$ active flavors. We
follow the definitions and conventions from
Ref.~\cite{Czakon:2008ii}. The relation between the coupling running
with $\nl$ and $\nl+1$ flavors can also be found there.

The results in
Eqns.~(\ref{eq:qqbar-1-beta},\ref{eq:qqbar-8-beta},\ref{eq:gg-1-beta},\ref{eq:gg-8-beta})
are in agreement with the ones extracted in
Ref.\cite{Hagiwara:2008df} from calculations of quarkonium
production at hadron colliders \cite{Kuhn:1992qw,Petrelli:1997ge}.

For applications to soft gluon resummation the above results are
also needed in Mellin space. One can easily switch between the two
representations of the fixed order threshold expansion and the
relevant formulas can be found, for example, in
Ref.~\cite{Moch:2008qy}. To further simplify the expressions, we
have effectively absorbed the Euler constant into the soft function
by switching to a modified Mellin moment ${\overline N} =
N\exp(\gamma_{\rm E})$. After performing the Mellin transformation
we can extract the exact expressions for the hard coefficients
$\sigma^{\rm H}_{q{\bar q},{\bf I}}$ as defined in
Eq.(\ref{eq:sigma-one-or-eight}) by keeping only the
non-$\log({\overline N})$ terms. The results read
\begin{eqnarray}
\sigma^{\rm H}_{q{\bar q},{\bf 1}} &=& {\cal O}(\as^2)
\, , \label{eq:qqbar-1-H}\\
\nonumber\\
\sigma^{\rm H}_{q{\bar q},{\bf 8}} &=& 1 + {\as\over \pi}\left[ C_F
\Bigg( - 8 + {2\pi^2\over 3} +3 \log 2 \Bigg) + C_A \left( {59\over
9} - {\pi^2\over 4} - 3 \log 2 \right) \right.\nonumber\\
&+& \left. \nl \left( -{5\over 9} + {2 \log 2\over 3} \right) -
{8\over 9} + \LOGmu2Om2 \left( - {3\over 2}C_F + {11\over 6} C_A -
{\nl + 1\over 3} \right) \right] + {\cal O}(\as^2)\, , \label{eq:qqbar-8-H}\\
\nonumber\\
\sigma^{\rm H}_{gg,{\bf 1}} &=& 1 + {\as\over \pi}\left[ C_F
\left(-5 + {\pi^2\over 4}\right) + C_A \left( 1 + {5\pi^2\over 12}
\right) - {1\over 3} \LOGmu2Om2 \right] + {\cal O}(\as^2) \, ,
\label{eq:gg-1-H}\\
\nonumber\\
\sigma^{\rm H}_{gg,{\bf 8}} &=& 1 + {\as\over \pi}\left[ C_F
\left(-5 + {\pi^2\over 4}\right) + C_A \left( 3 + {7\pi^2\over 24}
\right) - {1\over 3} \LOGmu2Om2 \right] + {\cal O}(\as^2) \, .
\label{eq:gg-8-H}
\end{eqnarray}

The coefficients
Eq.~(\ref{eq:qqbar-1-H},\ref{eq:qqbar-8-H},\ref{eq:gg-1-H},\ref{eq:gg-8-H})
are the main result from the present work. Due to the lack of a
singlet contribution through that order in perturbation theory (see
however the discussion at the end of Sec.~\ref{sec:NNLO}),
$\sigma^{\rm H}_{q{\bar q},{\bf 8}}$ coincides with the known
expressions for the color averaged cross section in the literature
\cite{Nason:1987xz,Beenakker:1988bq,Czakon:2008ii}. On the other
hand, the coefficients for the $gg$ reaction are new. They disagree
with the corresponding coefficients presented in
Ref.~\cite{Moch:2008qy}. Such disagreement is not surprising given
the fact that the color dependence of the hard coefficients has not
been taken into account in the attempt made in that reference to
exponentiate the NNLL soft logs.

Next we comment on the properties of the results above. The
vanishing of the LO singlet contribution in the $q{\bar q}$ is well
known, and is due to the fact that the LO reaction is mediated by an
s-channel gluon. Since the vanishing is not due to kinematics but
due to color effects, one expects that at higher orders this
property may no longer be true. At NLO the virtual corrections are
again zero due to the sandwiching of the one-loop amplitude with the
projected tree-level diagram. On the other hand, the square of the
one-gluon real emission diagrams is not identically zero. Our direct
calculation establishes that the color singlet contribution in the
$q{\bar q}$ reaction at NLO is suppressed by a factor of $\beta^2$
relative to the color averaged tree-level contributions and is thus
subleading. Such leading behavior is due to the absence of Coulomb
singularities and stronger suppression from the three-particle phase
space.

A rather striking feature of the exact color coefficients is that
their color dependence is ``standard'', i.e. they are simply
polynomials in $C_F$, $C_A$, etc. This is to be contrasted to the
color averaged coefficients used in the earlier literature where
color factors $\sim 1/(N_c^2-2)$ appear.

It is very interesting to try to estimate the size of the numerical
effect of the new terms in the $gg$ reaction derived here and in
Ref.~\cite{Czakon:2008ii}. To that end we calculate the hard
corrections $\sigma^{\rm H}_{ij,{\bf I}}$ and compare them to their
counterparts from Ref.~\cite{Bonciani:1998vc}. In the calculation we
take $N_c=3, ~ \mu^2=m^2=m_{\rm top}^2, ~ \as\left(\nf=\nl+1\right)
\approx 0.108$ and we restore the dependence of $\gamma_E$ as
explained above. We get the following results
\begin{eqnarray}
\sigma^{\rm H~ (BCMN)}_{gg} &=& 1 + {\as\over \pi}~ 14.39 ~ + {\cal
O}(\as^2) \, ,
\label{eq:num:BCMN}\\
\sigma^{\rm H~ (BCMN)}_{gg}\vert_{C_3~{\rm exact}} &=& 1 + {\as\over
\pi}~ 12.04 ~ + {\cal O}(\as^2) \, ,
\label{eq:num:BCMN-C3-mod}\\
\sigma^{\rm H}_{gg,{\bf 1}} &=& 1 + {\as\over \pi}~ 9.16 ~ + {\cal
O}(\as^2) \, ,
\label{eq:num:H-gg-1}\\
\sigma^{\rm H}_{gg,{\bf 8}} &=& 1 + {\as\over \pi}~ 13.19 ~ + {\cal
O}(\as^2) \, , \label{eq:num:H-gg-8}
\end{eqnarray}

The function $\sigma^{\rm H~ (BCMN)}_{gg}$ is defined as in
Ref.~\cite{Bonciani:1998vc}. The function $\sigma^{\rm H~
(BCMN)}_{gg}\vert_{C_3~{\rm exact}}$ has the same functional form as
the one in Eq.~(\ref{eq:num:BCMN}), but the value of the constant
$C_3$ has been modified from $C_3=37.23$ (as derived from
Ref.~\cite{Nason:1987xz} and as applied in
Ref.~\cite{Bonciani:1998vc}) to its exact value $C_3=34.88$ (as
derived in Ref.~\cite{Czakon:2008ii}).

Dividing Eq.~(\ref{eq:num:H-gg-1},\ref{eq:num:H-gg-8}) by
Eq.~(\ref{eq:num:BCMN}) and recalling
Eq.~(\ref{eq:sigma-one-or-eight}), we see that the effect of the
color dependence in the hard coefficients is a decrease of the
singlet (resp. octet) Sudakov cross section by 12\% (resp. 3\%)
compared to the ones in Ref.~\cite{Bonciani:1998vc}.

This is a large effect. Indeed, the shifts we observe in the Sudakov
factor are as large in size as the present conservative estimate
\cite{Cacciari:2008zb} of the theoretical uncertainty of the total
top pair production cross section and are significantly larger than the total
cross section uncertainty estimate in Ref.~\cite{Moch:2008qy}. A
detailed phenomenological investigation of the results derived here
will require a dedicated analysis.

It is also interesting to demonstrate the impact purely due to the
numerical uncertainty in the constant $C_3$. To that end we consider
the shift with respect to the results in
Ref.~\cite{Bonciani:1998vc}. Dividing Eq.~(\ref{eq:num:BCMN-C3-mod})
by Eq.~(\ref{eq:num:BCMN}) one can easily see that its effect is to
decrease the hard function $\sigma^{\rm H}_{gg}$ (and thus the
whole Sudakov cross section) by 5\%. This is also a very significant
effect given that its origin is pure numerics.

Another view of the impact of our results can be obtained by looking
at the cross section expanded to NNLO, similarly to what has been
done in \cite{Moch:2008qy}. Ignoring terms coming from Coulomb
enhancement, and cutting the logarithmic expansion at $\log^2 \beta$
(see discussion in Section~\ref{sec:NNLO}), the result presented in
Eq.~(21) of Ref.~\cite{Moch:2008qy} for the NNLO contribution reads
\begin{equation}
\label{eq:21MU} \sigma^{(2)}_{gg} = \sigma^{\rm
Born}_{gg}(\beta) \left(
   4608 \log^4 \beta
 + 1894.9 \log^3 \beta
 - 3.4811 \log^2 \beta + {\cal O}(\log \beta) \right) \; ,
\end{equation}
where the expansion parameter has been taken to be $\alpha_s/(4\pi)$
\begin{equation}
\sigma_{gg}(\beta) = \sigma^{\rm Born}_{gg}(\beta) +
\frac{\alpha_s}{4\pi} \sigma^{(1)}_{gg} + \left(
\frac{\alpha_s}{4\pi} \right)^2 \sigma^{(2)}_{gg} + {\cal
O}(\alpha_s^3) \; .
\end{equation}
It turns out that the coefficient of $\log^2 \beta$ exhibits an
accidental cancellation and is in fact given by
\begin{equation}
-14306.9505 + 384 C_3 \; .
\end{equation}
Inserting the exact value of $C_3$ derived in \cite{Czakon:2008ii},
the coefficient of $\log^2\beta$ in Eq.~(\ref{eq:21MU}) changes to
\begin{equation}
-912.35 \; ,
\end{equation}
i.e. the change of only 7\% in the value of $C_3$ results in a
magnification by a factor of about 260 of the coefficient of the
quadratic log of $\sigma_{gg}(\beta)$ at NNLO.

%
%
\section{Summary and Implications Beyond NLO/NLL} \label{sec:NNLO}

In this work we demonstrate that separate color singlet / color octet
hard (matching) coefficients need to be introduced in the Sudakov
total top quark pair production cross section. With the help of a
dedicated fixed order calculation we derive these coefficients in
analytic form. The difference between the hard coefficients for the two color
states has not been considered in earlier soft-gluon resummation
literature. We estimate the effect of these new 
contributions showing that they decrease the Sudakov total
cross section by 12\% in the singlet and by 3\% in the octet
channel. These shifts are large when compared to the current
conservative estimate of the uncertainty on the total top quark pair
production cross section.

The effect of these new contributions on the total top production
cross section will be somewhat reduced due to the subtraction of the ${\cal
  O}(\as)$ contributions from the Sudakov cross section (see
Ref.~\cite{Bonciani:1998vc} for detailed description of the NLO/NLL matching
procedure). However, we have also demonstrated that the new
corrections modify the terms in the Sudakov cross section at order ${\cal
  O}(\as^2)$ by a large amount. Therefore, a few percent effect on the total top
production cross section at the LHC can easily be anticipated. The exact
size of the impact of our findings can only be obtained from a
detailed phenomenological analysis.

With the results derived here and in Ref.~\cite{Czakon:2008ii} the
NLO/NLL program for the total top quark pair production cross
section at hadron colliders is now completed. The precision
requirements for this crucial observable for the LHC program are
very high and mandate improved theoretical precision. In principle,
to achieve that one has to go beyond the current NLO/NLL level of
accuracy.

Significant progress has already been made towards the direct fixed
order calculation of the top-pair production cross section at NNLO
\cite{Czakon:2007wk,Czakon:2007ej,Czakon:2008zk,Korner:2008bn,Kniehl:2008fd,Bonciani:2008az,Anastasiou:2008vd,Czakon:2008ii}.
In the near future, improvements can be expected from the careful
analysis of the new results reported here and in
Ref.~\cite{Czakon:2008ii}. The natural step beyond that is to try to
promote the resummation formalism to the NNLL level. We discuss in
the following how this can be done. Before we proceed, we
would like to make a comment concerning Ref.~\cite{Moch:2008qy}
where such an attempt has already been made. The comparison with the
direct exact calculation reported here shows that the one-loop hard
coefficients used in that reference are incorrect. Since this
discrepancy starts at the level of NLO/NLL it will clearly also
affect their predictions for the NNLL terms. For that reason we have
excluded the single log terms from our discussion around
Eq.~(\ref{eq:21MU}).

As we emphasized in our previous discussion, the basis
\cite{Kidonakis:1997gm} of the threshold exponentiation for the top
pair cross section is the singlet/octed diagonalization of the soft
anomalous dimension matrix in this special kinematics. Therefore,
the extension of the resummation formalism to the NNLL level,
requires to first verify if the corresponding two-loop {\it
massive} anomalous dimension diagonalizes in a similar manner. As of
writing of this paper there exists no such indication in the
literature. In fact, the only known \cite{Mitov:2006xs} property of
the two-loop massive anomalous dimension matrix is that it should
differ from the corresponding massless one (known through two-loops
from Ref.\cite{hep-ph/0606254,hep-ph/0607309}) by terms vanishing in
the massless limit as powers of the mass. Clearly this information
is insufficient to determine its behavior near threshold.

The next open problem in the NNLL resummation program would then be
the derivation of the two-loop anomalous dimension
$D_{ij,\bf 8}$ appearing in Eq.~(\ref{eq:Sudakov-exp}). Arguments about
its value are given in Ref.~\cite{Moch:2008qy}. In the present work,
based on an NLO fixed order calculation, we make no statement about it. That
would be a subject for future investigation.

The contributions from Coulomb singularities through two-loops are
known from other processes, and have been summarized in
Ref.~\cite{Moch:2008qy}.

Finally, we turn our attention to the hard coefficients $\sigma^{\rm
H}_{ij,{\bf I}}$ from Eq.~(\ref{eq:sigma-one-or-eight}). The
complete set at NLO has been presented in this work. The
corresponding two-loop corrections can only be extracted from a
future two-loop calculation of the top-production cross-section near
threshold. Clearly, this is a very demanding task. Moreover, there
might be an a priori non-vanishing contribution from the square of
the one-loop virtual diagrams to $\sigma_{q{\bar q},{\bf 1}}$
starting from order $\as^4$ (due to Coulomb enhancements and weaker
suppression from the two-particle phase space). If indeed nonzero,
it will contribute to a tower of NNLL soft logs and might have a
numerical impact on the Tevatron predictions. Such a possibility has
not been investigated so far in the literature.

As a final comment we would like to point to our discussion in
Ref.~\cite{Czakon:2008ii}, where we have argued about the rather
limited phenomenological value of truncating the all order
exponentiation to derive partial NNLO (or higher) terms. Specific
examples can be found in Fig.3 of Ref.~\cite{Melnikov:2005bx} and
Ref.~\cite{Czakon:2008ii}.

%
%
{\bf{Acknowledgments:}}
The research of A.M. is supported by a fellowship from the {\it US
LHC Theory Initiative} through NSF grant {\it PHY-0653342}. The work
of M.C. was supported in part by the Sofja Kovalevskaja Award of the
Alexander von Humboldt Foundation and by the ToK Program {\it
ALGOTOOLS} (MTKD-CD-2004-014319). A.M. would like to thank
G.~Sterman for very stimulating discussions. A.M. would also like to
thank M.~Cacciari and P.~Nason for valuable communications. Both
authors would like to thank DESY, Zeuthen for providing them with
access to their computing center.

%
%
{\footnotesize

}

\end{document}